\documentclass[12pt]{article}
\usepackage{amsmath}
\usepackage{amsfonts}
\usepackage{amssymb}
\usepackage{xcolor}
\usepackage{color}
\usepackage[english]{babel}
\usepackage{graphics}
\usepackage{graphicx}
\begin{document}
\title{Properties of the false vacuum \\ as the quantum unstable state}
\author{K. Urbanowski\footnote{e--mail:  K.Urbanowski@if.uz.zgora.pl, k.a.urbanowski@gmail.com}, \\
\hfill\\
University of Zielona G\'{o}ra, Institute of Physics, \\
ul. Prof. Z. Szafrana 4a, 65--516 Zielona G\'{o}ra, Poland.}
\maketitle
\begin{abstract}
We analyze properties of  unstable
vacuum states from the point of view of the quantum theory.
In the literature one can find some suggestions that
some of false (unstable) vacuum
states may survive up to times when their survival probability
has a non-exponential form. At asymptotically late times
the survival probability as a function of time $t$ has an
inverse power--like form.  We show that at this time region
the energy of the false vacuum states tends
to the energy of the true vacuum state as $1/t^{2}$ for
$t \to \infty$. This means that
the energy density in the unstable vacuum state should have analogous
properties and hence the  cosmological constant $\Lambda =
\Lambda (t)$ too. The conclusion is that
$\Lambda$ in the Universe with the unstable vacuum
should have a form of the sum of the "bare" cosmological constant and of the
term of  a type $1/t^{2}$:
$\Lambda(t) \equiv \Lambda_{bare} + d/ t^{2}$
(where $\Lambda_{bare}$ is the cosmological
constant for the Universe with the true vacuum).
\end{abstract}

\section{Introduction}

Broad discussion of the problem of false vacuum began after the publication of
pioneer papers by Coleman and his colleagues \cite{coleman1,callan,coleman2}.
The authors of these papers discussed
the problem of the stability of a physical system in
a state which is not an absolute minimum of its energy
density, and which is separated from the minimum by
an effective potential barrier.
In mentioned papers it was shown
that even if the state of the early Universe is too
cold to activate a "{\it thermal}" transition (via thermal
fluctuations) to the lowest energy (i.e. {\it "true vacuum"}) state,
a quantum decay from the false vacuum to the true vacuum
may still be possible through a barrier penetration
via  quantum tunneling.
Some time ago it appeared that
the problem of  the decay of the false vacuum state
can have a possible relevance in the process of
tunneling among the many vacuum states of the string
landscape (a set of vacua in the low energy
approximation of string theory).
In such cases
the scalar field potential driving inflation has a multiple,
low--energy minima or \textit{"false vacuua"}.
In such an situation the absolute minimum of the energy density
is the \textit{"true vacuum"}.

The problem of tunneling cosmological states was studied in many papers.
Part of these studies  was focused on analysis of the role of the
Standard Model Higgs boson in the inflationary cosmology.
Results presented in the pioneering  paper \cite{bezrukov1} show
that inflation can be natural consequence of the Standard Model.
This idea was analyzed  in  many papers (see eg. \cite{bezrukov2,bezrukov3,barvinsky3,barvinsky4}).
An application of some results of studies of the tunneling cosmological
state  to the inflationary cosmology driven by the Higgs boson
can be found eg. in  \cite{barvinsky1,barvinsky2}.
The dependence of the inflationary scenario and the dynamics of the Universe
(including the  problem of the stability of the Standard Model vacuum) on the
mass of the Higgs boson  was discussed e.g.
in \cite{bezrukov1,bezrukov2,bezrukov3,barvinsky3,barvinsky4,barvinsky1,barvinsky2}.
The discovery of the Higgs--like resonance at {{125 --- 126}} GeV (see, eg.,
\cite{kobahidze,degrassi,elias,wei})
was also the additional
 cause of much  discussions about the stability  of the false vacuum.
In \cite{degrassi} assuming  the validity of
the Standard Model up to Planckian energies it was shown that
a Higgs mass {{$m_{h} < 126$}} GeV implies that the electroweak
vacuum  is a metastable state. This means that
not only inflationary scenario and dynamics of the early Universe depend on
Higgs boson mass but also the  stability of the vacuum
state of the Universe, which is the Standard Model Higgs vacuum.
Thus a discussion of Higgs
vacuum stability can not  concentrate only on the standard model of elementary particles
but  must be considered also in a
cosmological framework, especially when analyzing
the process  of tunneling among the many vacuum states of the string
landscape.

Krauss and Dent analyzing a false vacuum decay \cite{krauss,winitzki}
pointed out that in eternal inflation,
many false vacuum regions
can survive up to the times much later than times when the exponential decay law
holds. This effect has a simple explanation: It may occur
even though regions of false vacua by assumption
should  decay exponentially, gravitational effects
force space in a region that has not decayed yet
to grow exponentially fast.

The aim of this talk is to discuss properties of
the false vacuum state as an unstable state,
and to analyze the late time behavior
of the energy of the false vacuum states.

\section{Briefly about quantum unstable states}

If {{$|M\rangle$}} is an initial unstable
state then the survival probability,{{ ${\cal P}(t)$}}, equals
{{${\cal P}(t) = |a(t)|^{2}$}},
where {{$a(t)$}} is the survival amplitude,
$ a(t) = \langle M|M;t\rangle$, $a(0) = 1$,
and, {{$|M;t\rangle =
\exp\,[-itH]\,|M\rangle$}},
{{$H$}} is the total Hamiltonian of the
system under considerations. (We use {$\hbar = c = 1$} units).
The spectrum, $\sigma(H)$, of $H$
is assumed to be bounded from below, $\sigma(H) =[E_{min},\infty)$
and $E_{min} > -\infty$.

From basic principles of quantum theory it is known that the
amplitude $a(t)$, and thus the decay law {{${\cal P}(t)$}} of the
unstable state $|M\rangle$, are completely determined by the
density of the energy distribution function $\omega({\cal E})$ for the system
in this state
\begin{equation}
a(t) = \int_{Spec.(H)} \omega({ E})\;
e^{\textstyle{-\,i\,{ E}\,t}}\,d{ E},
\label{a-spec}
\end{equation}
where $\omega({E}) \geq 0\;\;{\rm for} \;\;E
\geq E_{min}\;\;  {\rm and} \;\;\;\;\omega ({ E}) = 0
\;\;\; {\rm for} \;\;\;E < E_{min}$.
From this last condition and from the Paley--Wiener
Theorem it follows that there must be \cite{khalfin}
$|a(t)| \; \geq \; A\,\exp\,[ - b \,t^{q}]$
for $|t| \rightarrow \infty$. Here $A > 0,\,b> 0$ and $ 0 < q < 1$.
This means that the decay law ${\cal P}(t)$ of unstable
states decaying in the vacuum can not be described by
an exponential function of time $t$ if time $t$ is suitably long, $t
\rightarrow \infty$, and that for these lengths of time ${\cal P}(t)$
tends to zero as {$t \rightarrow \infty$}  more slowly
than any exponential function of {$t$}.
The analysis of the models of
the decay processes shows that
${\cal P}(t) \simeq
\exp\,[- {\it\Gamma}_{M}^{0}t]$,
 (where ${\it\Gamma}_{M}^{0}$ is the decay rate of the state $|M \rangle$),
to a very high accuracy  at the canonical decay times $t$:
From $t$ suitably later than the initial instant $t_{0}$
up to
$ t \gg \tau_{M} = 1/{{\it\Gamma}_{M}^{0}}$, ($\tau_{M}$ is
a lifetime), and smaller than $t = T$, where $T$ is
the crossover time and it denotes the
time $t$ for which the non--exponential deviations of $a(t)$
begin to dominate.

In general, in the case of quasi--stationary (metastable) states
it is convenient to express $a(t)$ in the
following form: $a(t) = a_{c}(t) + a_{lt}(t)$,
where $a_{c}(t)$ is the exponential (canonical) part of $a(t)$, that is $a_{c}(t) \stackrel{\rm def}{=}
N\,\exp\,[-it(E_{M}^{0} - \frac{i}{2}\,{\it\Gamma}_{M}^{0})]$,
($E_{M}^{0}$ is the energy of the system in the state
$|M\rangle$ measured at the canonical decay times,
$N$ is the normalization constant), and $a_{lt}(t)$ is the
non--exponential late time part of $a(t)$). For times $t \sim \tau_{M}$:
$|a_{c}(t)| \gg |a_{lt}(t)|$.
The crossover time $T$
can be found by solving the following equation,
\begin{equation}
|a_{c}(t)|^{\,2} = |a_{lt}(t)|^{\,2}.
\end{equation}
The amplitude $a_{lt}(t)$ exhibits inverse
power--law behavior at the late time region: $t \gg T$.
The integral representation (\ref{a-spec})
of $a(t)$ means that $a(t)$ is
the Fourier transform of the energy distribution
function $\omega(E)$. Using this fact we can find
asymptotic form of $a(t)$ for $t \rightarrow \infty$.
Results are rigorous \cite{ku-1,ku-2}.
If to assume that
\begin{equation}
\omega ({ E}) = ( { E} - { E}_{min})^{\lambda}\;
\eta ({ E})\; \in \; L_{1}(-\infty, \infty),
\label{omega-eta}
\end{equation}
(where $0 \leq \lambda < 1$), and
$\eta (E_{min}) \stackrel{\rm def}{=}
\eta_{0} > 0$, and $\eta^{(k)}({ E}) = \frac{d}{dE}\,\eta(E)$,
($k= 0,1,\ldots, n$),
exist and they are continuous
in $[{E}_{min}, \infty)$, and  limits
$\lim_{ E \rightarrow {E}_{min}+}\;
\eta^{(k)}( E) \stackrel{\rm def}{=} \eta_{0}^{(k)}$ exist, and
$\lim_{{ E} \rightarrow \infty}\;
( { E} - { E}_{min})^{\lambda}\,\eta^{(k)}({ E}) = 0$
for all above mentioned {$k$}, then
one finds that \cite{ku-1},
\begin{eqnarray}
a(t) & \begin{array}{c}
          {} \\
          \sim \\
          \scriptstyle{t \rightarrow \infty}
        \end{array} &
        (-1)\,e^{\textstyle{-{i}{ E}_{min} t}}\;
        \Big[
        \Big(- \frac{i}{t}\Big)^{\lambda + 1}\;
        \Gamma(\lambda + 1)\;\eta_{0}\; \label{a-eta} \\
        && +\;\lambda\,\Big(- \frac{i}{t}\Big)^{\lambda + 2}
        \;\Gamma(\lambda + 2)\;\eta_{0}^{(1)}\;+\;\ldots
        \Big]
         = a_{lt}(t),
            \nonumber
\end{eqnarray}
where $\Gamma(x)$ is Euler's gamma function.

From (\ref{a-eta}) it is seen that asymptotically late
time behavior of the survival amplitude $a(t)$
depends rather weakly on a specific form
of the energy density $\omega(E)$. The same
concerns a decay curves ${\cal P}(t) = |a(t)|^{2}$.
A typical form of a decay curve, that is the dependence
on time {$t$} of ${\cal P}(t)$ when $t$ varies from
$t = t_{0} =0$ up to  $t \,> \,30 \,\tau_{M}$ is
presented in Panels ${\bf\it A} $ of Figs. \ref{s10} and \ref{s50}.
Results presented in  these Figures
were obtained for
the  Breit--Wigner   energy distribution function,
$\omega ({E}) \equiv \omega_{BW} =
 \frac{N}{2\pi}\,  {\it\Theta} ( E - E_{min}) \
\frac{{\it\Gamma}_{M}^{0}}{({ E}-{ E}^{0}_{M})^{2} +
({\it\Gamma}_{M}^{0} / {2})^{2}}$, which corresponds
with $\lambda = 0$ in (\ref{omega-eta}).

\section{Instantaneous energy of the system in the unstable state}

The amplitude $a(t)$ contains information about
the decay law ${\cal P}(t)$ of the state $|M\rangle$, that
is about the decay rate ${\it\Gamma}_{M}^{0}$ of this state, as well
as the energy ${E}_{M}^{0}$ of the system in this state.
This information can be extracted from $a(t)$. Indeed if
$|M\rangle$ is an unstable (a quasi--stationary) state then
$a(t)  \cong N \exp\,[ - {i}({ E}_{M}^{0} -
\frac{i}{2}{\it\Gamma}_{M}^{0})\,t ]\,
= \,a_{c}(t)$ for {$t \sim \tau_{M}$.
So, there is
\begin{equation}
{E}_{M}^{0} - \frac{i}{2} {\it\Gamma}_{M}^{0} \equiv i
\,\frac{\partial a_{c}(t)}{\partial t} \;
\frac{1}{a_{c}(t)}, \;\;(t \sim \tau_{M}),
\label{E-iG}
\end{equation}
in the case of quasi--stationary states.

The standard interpretation and understanding of the quantum theory
and the related construction of our measuring devices are such that
detecting the energy ${E}_{M}^{0}$ and decay rate
${\it\Gamma}_{M}^{0}$ one is sure that the amplitude $a(t)$ has the
canonical form $a_{c}(t)$
and thus that the relation (\ref{E-iG})
occurs. Taking the above into account one can define the "effective
Hamiltonian", $h_{M}$, for the one--dimensional subspace of
states ${\cal H}_{||}$ spanned by the normalized vector
$|M\rangle$ as follows \cite{ku-1,ku-2,giraldi}
\begin{equation}
h_{M} \stackrel{\rm def}{=}  i \, \frac{\partial
a(t)}{\partial t} \; \frac{1}{a(t)}\;
\stackrel{\rm def}{=} {\cal E}_{M}(t)\,-\,\frac{i}{2}\,
\gamma_{M}(t). \label{h}
\end{equation}

In general, $h_{M}$ can depend on time $t$, $h_{M}\equiv
h_{M}(t)$. One meets this effective Hamiltonian when one starts
with the Schr\"{o}dinger Equation
for the total state
space ${\cal H}$ and looks for the rigorous evolution equation for
the distinguished subspace of states ${\cal H}_{||} \subset {\cal
H}$ (see Appendix and also \cite{ku-1,ku-2}).
Using $h_{M}(t)$
one finds the following expressions for the
energy and the decay rate of the system in the state $|M\rangle$
under considerations, to be more precise for
the instantaneous energy ${\cal E}_{M}(t)$ and the instantaneous decay rate,
$\gamma_{M}(t)$ (see Appendix and \cite{ku-1}),
\begin{equation}
{\cal E}_{M}\equiv {\cal E}_{M}(t) = \Re\,[h_{M}(t)], \;\;\;\;\;
\gamma_{M} \equiv \gamma_{M}(t) = -\,2\,\Im\,[h_{M}(t)],\label{E(t)}
\end{equation}
where $\Re\,(z)$ and $\Im\,(z)$ denote the real and imaginary parts
of $z$ respectively.

Defining
\begin{equation}
h_{M}^{lt}(t) \stackrel{\rm def}{=} {h_{M}(t)\vline}_{\,t \rightarrow \infty}
\label{h-lt}
\end{equation}
one can calculate the asymptotic late time
form $h_{M}^{lt}(t)$ of $h_{M}(t)$ using asymptotic late time form
of the amplitude $a_{lt}$(t):
\begin{equation}
h_{M}^{lt}(t) = i \, \frac{\partial a_{lt}(t)}{\partial t} \; \frac{1}{a_{lt}(t)}.
\label{h-lt-1}
\end{equation}
So, starting from the  asymptotic
expression (\ref{a-eta}) for $a(t)$ and using (\ref{h-lt-1})
 one can find   $h_{M}^{lt}(t)$ for  times
 $t \gg T$ as a function of a small parameter
$x=1/t$: $h_{M}^{lt}(x) = \frac{b(x)}{a_{lt}(x)}$,
where $b(x) = i {\frac{\partial a_{lt}(t)}{\partial t}  \vline}_{x=\frac{1}{t}}$,
 then expanding such obtained $h_{M}^{lt}(x)$ in Taylor series about $x=0$
one obtains after some algebra that
\begin{equation}
{h_{M}(t)\vline}_{\,t \rightarrow \infty} = h_{M}^{lt}(t)
\simeq { E}_{min} + (-\,\frac{i}{t})\,c_{1} \,
+\,(-\,\frac{i }{t})^{2}\,c_{2} \,+\,\ldots, \label{h-infty-gen}
\end{equation}
where $ c_{i} = c_{i}^{\ast},\;\;i = 1,2,\ldots$;
(coefficients {$c_{i}$} depend on  {$\omega (E)$}).
This  last relation means that
\begin{equation}
{\cal E}_{M}(t) \,\simeq\,  E_{min} \, -
\,\frac{c_{2}}{t^{2}} \ldots, \;\;\;\;\;\;\;
\gamma_{M}(t)\, \simeq \, 2\,\frac{c_{1}}{t} \,+\ldots, \;\;\;({\rm for}
\;\;t \gg T). \label{E(t)+G(t)}
\end{equation}
These properties take place for  all unstable states
which survived up to times $t \gg T$.
Note that from (\ref{E(t)+G(t)}) it follows that
$\lim_{t \rightarrow \infty}\, {\cal E}_{M}(t) = E_{min}$
and $\lim_{t \rightarrow \infty}\,\gamma_{M}(t) =0$.

For the density {$\omega(E)$} of the form (\ref{omega-eta})
(i. e. for $a(t)$ having the asymptotic
form given by (\ref{a-eta}))  we have
\begin{equation}
c_{1} = \lambda + 1, \;\;\;\;c_{2} =
(\lambda + 1)\,\frac{\eta^{(1)}(E_{min})}{\eta (E_{min})}.
\label{c-i}
\end{equation}
The energy distribution densities $\omega (E)$ considered
in quantum mechanics and in quantum field theory can be described by
$\omega (E)$ of the form  (\ref{omega-eta}), eg.  quantum
field theory models analyzing two particle decays
correspond with $\lambda = \frac{1}{2}$.

A general form of
\begin{equation}
\kappa (t) = \frac{{\cal E}_{M}(t) - E_{min}}{E_{M}^{0} - E_{min}}, \label{kappa}
\end{equation}
 as a function of time $t$ varying from $t = t_{0} =0$
 up to $ t > T$ is presented
in  Panels ${\bf\it B}$ of Figs. \ref{s10} and \ref{s50}.
These results were obtained for  $\omega (E) = \omega_{BW}(E)$.
The crossover time {$T$}, that is the time region where fluctuations
of ${\cal P}(t)$ and ${\cal E}_{M}(t)$ take place depends on the
value of the parameter $s_{0} = (E_{M}^{0} - E_{min})/{\it\Gamma}_{M}^{0}$
in the model considered: The smaller $s_{0}$ the shorter $T$.

\begin{figure}[h!]
\begin{center}
\includegraphics[width=100mm]{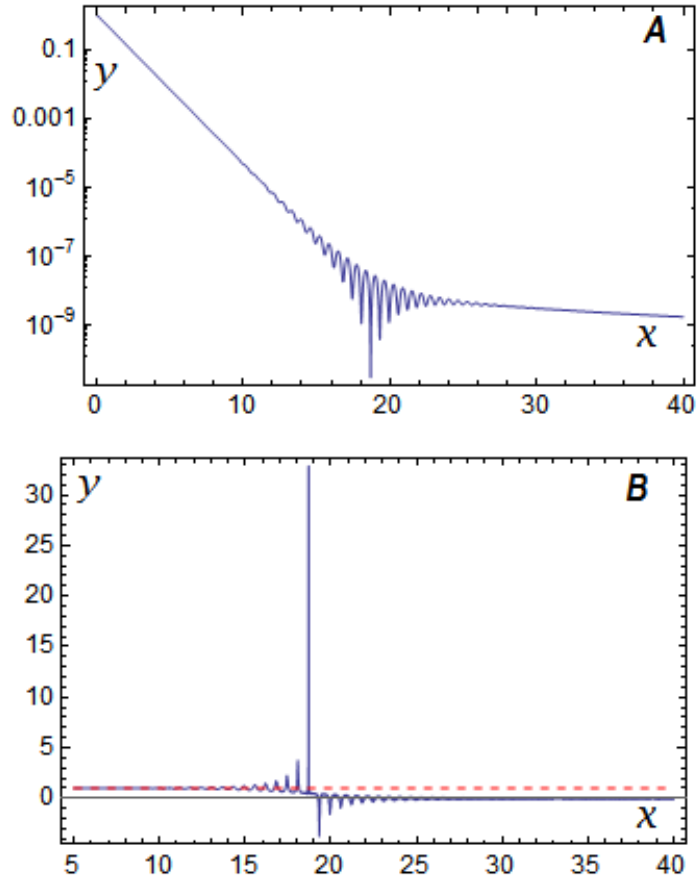}
\end{center}
\caption{ The case $s_{0} =10$. Panel ${\bf\it A}$ --- Axes:
$y= {\cal P}(t)$, (The logarithmic  scale),
$x =t / \tau_{M}$.  Panel ${\bf\it B}$, Axes: $y = \kappa (t)$, $ x=t/\tau_{M}$.
The horizontal red dashed line denotes $y = \kappa (t) = 1$.}
\label{s10}
\end{figure}

\begin{figure}
\begin{center}
\includegraphics[width=100mm]{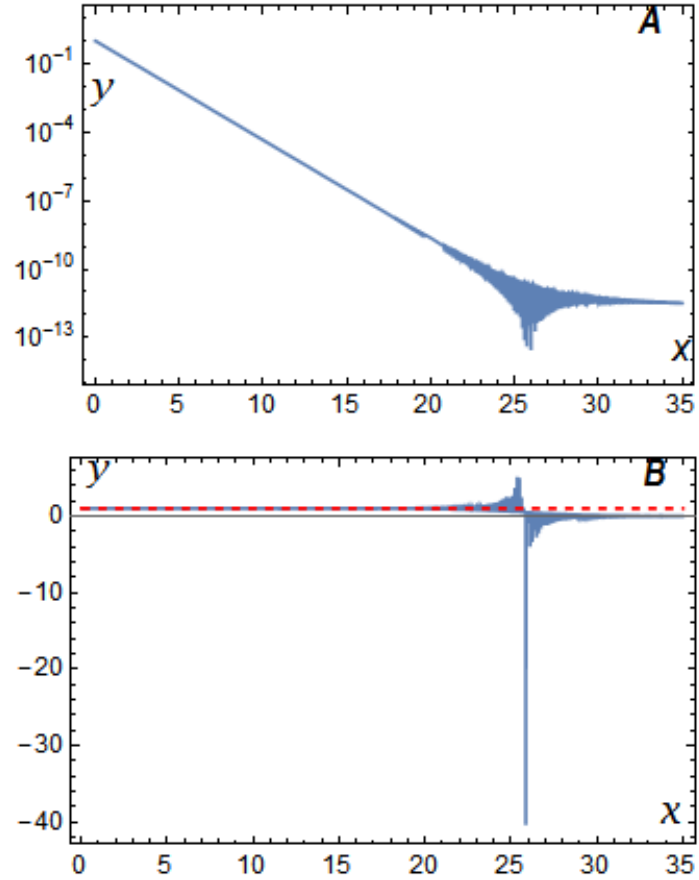}
\end{center}
\caption{  The case $s_{0} =50$. Panel ${\bf\it A}$ --- Axes:
$y= {\cal P}(t)$, (The logarithmic scale), $x =t / \tau_{M}$.
Panel ${\bf\it B}$, Axes: $y = \kappa (t)$, $ x=t/\tau_{M}$. The
horizontal red dashed line denotes $y = \kappa (t) = 1$,
that is ${\cal E}_{M}(t) = E_{M}^{0}$.}
\label{s50}
\end{figure}

The  equivalent formula for $h_{M}(t)$ is given by the following relation:
\begin{equation}
h_{M} (t) \equiv \frac{\langle M|H|M (t)\rangle}{\langle M |M (t)\rangle}.
\label{h-equiv}
\end{equation}
This last relation
explains  properties of ${\cal E}_{M}(t)$ at different
time regions which one can see in Panels $B$ of
Figs. \ref{s10} and \ref{s50}. Indeed, if to rewrite the numerator
of the righthand side of (\ref{h-equiv}) as follows,
\begin{equation}
\langle M|H|M (t)\rangle \equiv \langle
M|H|M\rangle\,a(t)\,+\,\langle M |H|M (t)\rangle_{\perp}, \label{perp}
\end{equation}
where $|M (t)\rangle_{\perp} = Q|M (t)\rangle$,
$Q = \mathbb{I} - P$ is the projector onto the subspace od decay products,
$P = |M\rangle\langle M|$ and $\langle M|M (t)\rangle_{\perp} = 0$,
then one can see
that there is a permanent contribution of
decay products described by $|M (t)\rangle_{\perp}$
to the energy of the unstable state considered.
The intensity of this contribution depends on time $t$.
This contribution into the instantaneous energy
is practically negligible small and constant
in time at  canonical decay times
whereas at the transition times, when $t \sim T$, it
is fluctuating function of time and the amplitude
of these fluctuations may be significant. What is more
relations (\ref{h-equiv}) and (\ref{perp}) allow
one to proof that in the case of unstable
states $\Re\,[h_{M}(t)] \neq const$. Using these relations
one obtains that
\begin{equation}
h_{M}(t) = E_{M} +\,
\frac{\langle M |H|M (t)\rangle_{\perp}}{a(t)}, \label{h-perp-1}
\end{equation}
where
 $E_{M}$ is the expectation value of $H$: $E_{M} = \langle M|H|M\rangle$.
From this relation one can see that
$h_{M}(0) = E_{M}$ if the matrix elements
$\langle M|H|M\rangle$ exists. It is because \linebreak
$|M (t=0)\rangle_{\perp} =0$ and $a(t=0)=1$.
Now if to assume that for $0\leq t_{1} \neq t_{2}$ there
is $\Re\,[h_{M}(0)] = \Re\,[h_{M}(t_{1})] = \Re\,[h_{M}(t_{2})] = const$
then one immediately
conclude that there should be $\Re\,[h_{M}(t)] = E_{M}$
 for any $t \geq 0$. Unfortunately
such an observation contradicts implications of  (\ref{h-perp-1}):
From this relation it follows that $\Re\,[\frac{\langle M |H|M (t)\rangle_{\perp}}{a(t)}]
\neq 0$ for $t > 0$ and thus $\Re\,[h_{M}(t > 0)] \neq E_{M}
 \equiv \Re\,[h_{M}(0)]$ which shows that
$\Re\,[h_{M}(t)] \equiv {\cal E}_{M}(t)$ can not be constant in time.

\section{Connections with the cosmology}

From the point of the quantum theory the decay of the false vacuum  is the
 quantum decay process \cite{coleman1,callan,coleman2,krauss,winitzki}.
What is more some cosmological scenario
predict the possibility of decay of the Standard Model vacuum at an inflationary stage of
the evolution of the universe (see eg. \cite{branchina} and also \cite{bezrukov4} and reference therein). Of course
this decaying  Standard Model vacuum is described by the quantum state corresponding to a local minimum
of the energy density which is not the absolute minimum of the energy density
of the system considered.
This means that  state vector corresponding  to the false vacuum is
a quantum unstable (or metastable) state. Therefore
 all the general properties of quantum unstable states  must also occur
 in the case of such a quantum unstable state as false vacuum.
This applies in particular to such properties as late time
deviations from the exponential decay law
and  properties of the energy $ {\cal E}^{false}_{0}(t)$  of the
system in the quantum false vacuum  state at late times $t > T$.

The cosmological scenario in which false vacuum may decay at the inflationary stage of the universe corresponds
with the hypothesis analyzed by Krauss and Dent in \cite{krauss}.
Namely in the mentioned paper the hypothesis
that some false vacuum regions do survive well
up to the time $T$ or  later was formulated.
So, let $|M\rangle = | 0\rangle^{false}$,
be a false, $|0\rangle^{true}$ -- a true, vacuum
states and   $E^{false}_{0}$ be the energy of a state
corresponding to the false vacuum measured at the canonical decay time
and  $E^{true}_{0}$ be the energy of true vacuum (i.e.
the true ground state of the system).
As it is seen from the results presented in previous
Section, the problem is that the energy of those false
vacuum regions which survived up to $T$ and much later
differs from $E^{false}_{0}$  \cite{ku-3}.

Now, if  one assumes that $E^{true}_{0} \equiv E_{min}$ and
$E_{0}^{false} =  E_{M}^{0}$ and takes into account results
of the previous Section (including those in Panels $\bf\it B$
of Figs. \ref{s10} and  \ref{s50})
then one can conclude that the energy of the system in  the
false vacuum state has the following general properties:
\begin{equation}
{\cal E}^{false}_{0}(t) = E^{true}_{0}  + \Delta E \;\cdot\;\kappa(t),
\label{E-false-(t)}
\end{equation}
where $\Delta E = E_{0}^{false} \,-\, E^{true}_{0}$
and $\kappa (t)
\simeq 1$ for $t \sim \tau_{0}^{false} < T$, (where $\tau_{0}^{false}$ is the life time of the unstable false vacuum state).
$\kappa (t)$ is a fluctuating function of $t$   at
$t \sim T$  and   $ \kappa (t) \propto \frac{1}{t^{2}}$
for $t \gg T$.

At asymptotically late times,
$t \gg T$, one finds that
\begin{equation}
{\cal E}^{false}_{0}(t) \simeq E^{true}_{0}  -
\frac{c_{2}}{t^{2}}\ldots \;\; \neq\,E^{false}_{0},
\label{E-false-infty}
\end{equation}
where $c_{2} = c_{2}^{\ast}$ and it can be
positive or negative depending on the model considered.
Similarly $\gamma^{false}_{0}(t) \simeq   + 2\,c_{1}/t\, \ldots$ for $t \gg T$.
Two last properties of the false vacuum states mean that
\begin{equation}
{\cal E}^{false}_{0}(t) \rightarrow E^{true}_{0} \;\;{\rm and}\;\;
\gamma^{false}_{0}(t) \rightarrow 0
\;\;{\rm as}\;\; t\rightarrow \infty. \label{E-false-lim}
\end{equation}

Now if one wants to generalize
the above results obtained on the basis of quantum
mechanics to quantum field theory
one should take into account  among others a volume factors
so that survival probabilities per unit volume per unit
time should be considered.
The standard false vacuum decay
calculations shows that the same volume factors should appear
in both early and late time decay rate estimations
(see Krauss and Dent \cite{krauss,ku-4}). This means that
the calculations of cross--over time $T$ can be
applied to survival probabilities per unit
volume.  For the same reasons  within the quantum field
theory the quantity  ${\cal E}_{M}(t)$ can be replaced by
the energy per unit volume $\rho_{M}(t) = {\cal E}_{M}(t)/V$
because these volume factors $V$  appear in the numerator
and denominator of the formula (\ref{h}) for $h_{M}(t)$.
This conclusion seems to hold when considering the
energy ${\cal E}_{0}^{false}(t)$ of the system in false
vacuum state $|0\rangle^{false}$ because Universe is
assumed to be homogeneous and isotropic at suitably
large scales. So at such scales to a sufficiently
good accuracy we can extract properties of the energy
density $\rho_{0}^{false}= E_{M}^{0}/V =
E_{0}^{false}/V$ of the system in the false
vacuum state $|0\rangle^{false}$ from properties of
the energy ${\cal E}_{0}^{false}(t)$ of the system
in this state defining $\rho_{0}^{false}(t)$ as
$\rho_{0}^{false}(t) = {\cal E}_{0}^{false}(t)/V$.
This means that in the case of
 a meta--stable (unstable or decaying, false) vacuum
the following important property of $\kappa (t)$ holds:
\begin{eqnarray*}
\kappa(t)
&\equiv &  \frac{\rho_{0}^{false}(t) - \rho_{bare}}{\rho_{0}^{false} - \rho_{bare}},
\end{eqnarray*}
where $\rho_{bare} = E_{min}/V$ is the energy density of the true (bare) vacuum.
From the last equation the following relation follows
\[
\rho_{0}^{false}(t) - \rho_{bare} = (\rho_{0}^{false} - \rho_{bare})\,\kappa (t).
\]
Thus, because  for $t < T$ there is $\kappa (t) \simeq 1$, one finds that
\[
\rho_{0}^{false} (t) \simeq \rho_{0}^{false},\;\;{\rm for}\;\; t <T,
\]
whereas for $t \gg T$ we have
\begin{equation}
\rho_{0}^{false}(t) - \rho_{bare} = (\rho_{0}^{false} -
\rho_{bare})\,\kappa (t) \simeq \pm \,d_{2}\,\frac{\hbar^{2}}{t^{2}},\;\; (t \gg T), \label{rho}
\end{equation}
where $d_{2}=d_{2}^{\ast}$. The units $\hbar = 1 = c$
will be used in the next formulas.
Analogous relations (with the same $\kappa (t))$ take place for
$\Lambda (t) = \frac{8\pi G}{c^{2}}\,\rho(t)$}, or
{$\Lambda (t) = 8\pi G\,\rho(t)$ in $\hbar = c =1$ units:
\begin{equation}
\Lambda (t) - \Lambda_{bare} = (\Lambda_{0} - \Lambda_{bare})\,\kappa (t), \label{lambda}
\end{equation}
or,
\begin{equation}
\Lambda (t) = \Lambda_{bare} +(\Lambda_{0} - \Lambda_{bare})\,\kappa (t). \label{lambda1}
\end{equation}

One may expect that $\Lambda_{0}$ equals to  the cosmological
constant calculated within quantum field theory. From (\ref{lambda1})
it is seen that for $t < T$,
\begin{equation}
 \Lambda (t) \simeq \Lambda_{0}, \;\;\;{\rm for}\;\;\;(t < T),
 \end{equation}
 because $\kappa (t < T) \simeq 1$. Now if to assume that $\Lambda_{0}$
 corresponds to the value of the cosmological "constant" $\Lambda$ calculated
 within the quantum field theory, than one should expect that \cite{sz1}
 \begin{equation}
 \frac{\Lambda_{o}}{\Lambda_{bare}} \geq 10^{120}, \label{lambda2}
 \end{equation}
 which allows one to write down Eq. (\ref{lambda1}) as follows
 \begin{equation}
 \Lambda(t) \simeq \Lambda_{bare} + \Lambda_{0}\,\kappa(t). \label{lambda3}
 \end{equation}
 Note that for $t \gg T$ there should be (see (\ref{rho}))
 \begin{equation}
\Lambda_{0}\, \kappa (t) \simeq {8\pi G}\, \frac{d_{2}}{t^{2}}
\equiv \pm \frac{\alpha^{2}}{t^{2}},\;\;{\rm for}\;\; ( t \gg T). \label{lambda4}
 \end{equation}

\section{ Final Remarks}

Parametrization following from quantum theoretical treatment of
meta\-stable vacuum states can explain why the cosmologies with
the time--depen\-dent cosmological constant $\Lambda (t)$ should
be considered. Such a parametrization may help to explain the
cosmological constant problem \cite{Weinberg,Caroll}. From the
literature we know that the time dependence of  $\Lambda$ of
the type $\Lambda (t) = \Lambda_{bare} + \frac{\alpha^{2}}{t^{2}}$
was considered by many authors:
Similar form of $\Lambda$ was obtained in \cite{Canuto}, where
the  invariance under scale transformations of
the generalized Einstein equations was studied.
Such a time dependence of $\Lambda$ was
postulated also in \cite{Lau} as the result of
the analysis of the large numbers hypothesis.
The cosmological model with  time dependent {$\Lambda$}
of the above postulated form was studied also in \cite{Berman}.
This form of $\Lambda$  was assumed in eg.
in \cite{Lopez} but there was no any explanation what
physics suggests such the choice. Cosmological model with
time dependent {$\Lambda$} were also studied in much more recent papers.

The advantage of the formalism presented in Sect. 4 is that
it takes into account that the decay proces of the metastable
vacuum state is the quantum decay process.
So in the case of the universe with metastable (false) vacuum
when one realizes that the decay of this unstable vacuum
state is the quantum decay process then it
emerges automatically that there have to exist the true
ground state of the system that is the true (or bare)  vacuum
with the minimal energy, $E_{min}> - \infty$, of the system
corresponding to him and equivalently, $\rho_{bare} = E_{min}/V$, or $\Lambda_{bare}$.
Also in this case such  $\Lambda \equiv
\Lambda(t)$ emerges   that at suitable late times it has
the form described by relations (\ref{lambda3}), (\ref{lambda4}).
In such a case   the function $\kappa(t)$ given by the relation
(\ref{kappa}) describes time dependence for all times {$t$} of the
energy density $\rho_{M}(t)$ or the cosmological "constant" $\Lambda_{M}(t)$
and it general form is presented in Panels $\bf\it B$ in Figs. \ref{s10} and \ref{s50}.
Note that results presented in Sections 2 --- 4 are rigorous.
The formalism mentioned was applied
in \cite{ku-4,sz1}, where cosmological models
with $\Lambda(t) = \Lambda_{bare} \pm \frac{\alpha^{2}}{t^{2}}$ were
studied: The nice and may be the most promising result is reported in \cite{sz1}
where using the parametrization following from the mentioned
quantum theoretical analysis of the decay process of the unstable
vacuum state an attempt was made to explain
the small today's value of the cosmological constant {$\Lambda$}.
So we can conclude that
formalism and the approach described in this paper
and in \cite{ku-4,sz1} is promising and
can help to solve the cosmological constant and
other cosmological problems and it needs further studies.
What is more, in the light of
the LHC result concerning the mass $m_{H}$ of the Higgs
boson  and cosmological consequences of this result
such conclusions seem to be reasonable and justified.
It is because
according to the observation  that result
$m_{H} < 126$ GeV  may impliy instability of the electroweak vacuum
and that there are cosmological scenario that
predict  even the possibility of decay of this vacuum at an inflationary stage of
the evolution of the universe \cite{bezrukov4,branchina}.\\
\hfill\\
\noindent
\textbf{Acknowledgments:} The work was supported in part by the NCN grant No
DEC-2013/09/B/ST2/03455.\\
\hfill\\

\hfill\\
{\large \textbf{Appendix}}\\

Let us consider general properties of $h_{M}(t)$.
In the proper quantum mechanical calculations of the
decay processes one always starts from the Schr\"{o}dinger equation
 \begin{equation}
i \frac{\partial}{\partial t} |M (t) \rangle = H |M (t)\rangle.  \label{Schrod}
\end{equation}
 Here $|M (t)\rangle \in {\cal H}$,
 $H$ denotes the total self--adjoint Hamiltonian
 and $|M\rangle = |M (0)\rangle \in {\cal H}$ is the initial
 unstable state of the system. There is $\langle M|M\rangle = 1$.
To be more precisely the problem reduces to replacing Schr\"{o}\-din\-ger equation
(\ref{Schrod}) by two coupled equations and then by solving them to obtain  the equation for the amplitude $a(t)$.
Using projection operators defined in Sec. 3, $P = |M\rangle \langle M|$ and $Q = \mathbb{I} - P$
one can see that simply
\begin{equation}
P|M;t\rangle \equiv a(t)\,|M\rangle, \label{P-phi-t}
\end{equation}
and instead of (\ref{Schrod}) one obtains two coupled equations (see eg. \cite{pra})
\begin{eqnarray}
i\,\frac{\partial }{\partial t}\,P|M (t)\rangle &=& PHP|M (t)\rangle + PHQ|M (t)\rangle, \label{E-1} \\
i\,\frac{\partial }{\partial t}Q|M (t)\rangle &=& QHQ |M (t)\rangle + QHP|M (t)\rangle . \label{E-2} 
\end{eqnarray}
The initial conditions are:
\begin{equation}
P|M (0)\rangle = |M\rangle, \;\;\;\;Q|M (0)\rangle = 0. \label{P-Q-0}
\end{equation}
Taking into account initial conditions (\ref{P-Q-0})  and inserting  a
solution $Q|M (t)\rangle$ of Eq. (\ref{E-2}) into Eq. (\ref{E-1}) one obtains the equation for the amplitude $a(t)$:
\begin{equation}
\Big(i\frac{\partial}{\partial t} - PHP\Big)\,a(t)\,|M\rangle = - i \int_{0}^{\infty}\,k(t - \tau)\,a(\tau)\,|M\rangle\,d\tau,
\label{KR-1}
\end{equation}
where $t \geq 0$, and
\begin{equation}
k(t) = \Theta (t)\,\langle M|HQ\,e^{\textstyle{-itQHQ}}\,QH|M\rangle, \label{K(t)}
\end{equation}
($\Theta (t)$ is a unit step function). There is
$$
PHP \equiv \langle M|H|M\rangle\,|M \rangle \equiv E_{M} \,|M\rangle,
$$
where $E_{M}$.  This last property means that Eq. (\ref{KR-1}) can be rewritten as follows
(see \cite{pra} or \cite{peres} where similar, equivalent equations were considered),
\begin{equation}
\Big(i\frac{\partial}{\partial t} - E_{M}\Big)\,a(t) = - i \int_{0}^{\infty}\,k(t - \tau)\,a(\tau)\,d\tau.  \label{KR-2}
\end{equation}
Equations (\ref{KR-1}) and (\ref{KR-2}) are exact. Performing calculations of decay
processes one usually uses approximate methods to solve (\ref{KR-2}). In order to obtain
corrections to the energy of the unstable states it is convenient to replace integro--differential equation (\ref{KR-2}) by equivalent, only differential one:
\begin{equation}
\Big(i\frac{\partial}{\partial t} - E_{M}\,-\,v_{M}(t)\Big)\,a(t) \equiv  0, \label{E+v}
\end{equation}
where the "quasi--potential" $v_{M}(t)$ can be found by solving the non--linear equation:
\begin{equation}
v_{M}(t)\,a(t) =  - i \int_{0}^{\infty}\,k(t - \tau)\,a(\tau)\,d\tau. \label{v-1}
\end{equation}
The approximate solution  of this equation to lowest nontrivial order
can be found using the "free" solution $a^{(0)}(t)$ of (\ref{KR-2}), that is
using  solutions of the following equation
\begin{equation}
\Big(i\frac{\partial}{\partial t} - E_{M}\Big)\,a^{(0)}(t) = 0, \label{free}
\end{equation}
for the initial condition $a^{(0)}(0)=a(0) =1$. Inserting solution
 $a^{(0)}(t) = \exp\,[-itE_{M}]$ of (\ref{free}) into (\ref{v-1}) one obtains
(for details see \cite{pra}, Eq. (32) --- (35)),
\begin{equation}
v_{M}(t) \simeq  v_{M}^{(1)}(t) = -i \int_{0}^{\infty}\,k(t - \tau)\, e^{\textstyle{i(t-\tau)E_{M}}}\,d\tau. \label{v-3}
\end{equation}
This relation leads to the following expression for $v_{M}^{(1)}(t)$:
\begin{equation}
v_{M}^{(1)}(t) = \langle M|HQ\,\frac{e^{\textstyle{-it(QHQ - E_{M})}}\,-\,1}{QHQ\,-\,E_{M}}\,HQ|M\rangle \equiv
- \Delta^{(1)}_{M}(t) - \frac{i}{2}\,{\it\Gamma}^{(1)}_{M}(t), \label{v(1)}
\end{equation}
where $\Delta^{(1)}_{M}(t)$ and ${\it\Gamma}^{(1)}_{M}(t)$ are real.
For large $t$ this approximate result coincides with the Weisskopf--Wigner result \cite{WW,pra}:
\begin{equation}
\lim_{t\to\infty}\,v_{M}^{(1)}(t) = v_{M}^{(1)} = - \Delta^{(1)}_{M} - \frac{i}{2}{\it\Gamma}^{(1)}_{M}, \label{v-infty}
\end{equation}
where
\begin{equation}
\Delta^{(1)}_{M} = \langle M|HQ\;\,{\rm P}_{v}\,\frac{1}{ QHQ - E_{M} }\;\,QH|M\rangle, \label{Delta}
\end{equation}
(here ${\rm P}_{v}$ denotes principal value), and $\Delta^{(1)}_{M}$ is the correction to the energy of unstable state, and
\begin{equation}
{\it\Gamma}^{(1)}_{M} = 2\pi \langle M|HQ\,\delta(QHQ - E_{M})\,QH|M\rangle, \label{Gamma(1)}
\end{equation}
is the decay width.  More detailed considerations show that these approximate results
 describe behavior of the unstable $|M\rangle$ accurate enough only for canonical decay times
 (i.e. when the exponential decay law holds with sufficient accuracy \cite{khalfin} ).

Taking into account relations (\ref{E+v}), (\ref{v(1)}), (\ref{v-infty}) and (\ref{Delta})
we conclude that the energy $E_{M}^{0}$ of an unstable state  $|M\rangle$ at canonical decay times equals
\begin{equation}
E_{M}^{0} \simeq E_{M} - \Delta^{(1)}_{M} \equiv E_{M} + \Re\,[ v_{M}^{(1)}]. \label{E-phi}
\end{equation}
On the other hand from (\ref{E+v}) it follows that
\begin{equation}
E_{M} + v_{M}(t) \equiv \frac{i}{a(t)}\;\frac{\partial a(t)}{\partial t}. \label{E+v=h-1}
\end{equation}
This relation together with (\ref{h}) means that simply
\begin{equation}
h_{M}(t) \equiv E_{M} + v_{M}(t), \label{h=E+v}
\end{equation}
and Eq. (\ref{E+v}) can be written as
\begin{equation}
\Big(i\frac{\partial}{\partial t}\,-\,h_{M}(t)\Big)\,a(t) \equiv  0, \label{eq-for-h}
\end{equation}
where simply $h_{M}(t)$ is the effective hamiltonian governing time e
volution of the unstable state considered. Comparing (\ref{h=E+v}) and
(\ref{h-perp-1}) one finds that equivalently
\begin{equation}
v_{M}(t) \equiv  \frac{\langle M |H|M (t)\rangle_{\perp}}{a(t)}. \label{h-perp-2}
\end{equation}

From the above analysis it is seen that
\begin{equation}
v_{M}(t=0) =0, \label{v(0)}
\end{equation}
which means that $\Re\,[h_{M}(t=0)] =  E_{M} $.

At canonical decay times,
we have
\begin{eqnarray}
\Re\, [h_{M}(t)] &=& E^{0}_{M} \simeq E_{M} - \Delta^{(1)}_{M},
\;\;\;({\rm for}\;\;t \sim \tau_{M}), \label{h(tau)}  \\
\Im\,[h_{M}(t)] &=& -\,\frac{1}{2}\, {\it\Gamma}_{M}^{\,0} \simeq -\, \frac{1}{2}\,{{\it\Gamma}}^{(1)}_{M},
\;\;\;({\rm for}\;\;t \sim \tau_{M}).
\label{Gamma(tau)}
\end{eqnarray}
Therefore
when one measures the energy of the unstable state considered at canonical decay times, that is at
times  $t \sim \tau_{M}$
then one expects to obtain  the energy $E_{M}^{0}$ defined by formula
(\ref{h(tau)}) as the result of the measurement.

So, as it follows from relations
(\ref{E-phi}) --- (\ref{h(tau)})
the energy of the system in the unstable state $|M\rangle$, ${\cal E}_{M}(t)$
is equal to the real part of the effective hamiltonian $h_{M}(t)$:
\begin{equation}
{\cal E}_{M}(t) = \Re \, [h_{M}(t)] = E_{M} - \Delta_{M}(t), \label{E=Re-h}
\end{equation}
where $\Delta_{M}(t) \stackrel{\rm def}{=} \Re\,[v_{m}(t)]$ and $\Delta_{M}(0) = 0$ 
and the sign of $\Delta_{M}(t) $ depends on the model considered.
Strictly speaking, if to take into account  all properties of $\Re\,[h_{M}(t)]$, the energy ${\cal E}_{M}(t)$
is  the instantaneous energy of the system in the unstable state $|M\rangle$
and this energy  is equal to the sum of the expectation value, $E_{M}$ of
the total Hamiltonian and the contribution
 $\Delta_{M}(t)$ coming from the interactions responsible for decay and regeneration processes.
 To complete the above considerations one should mention connections of late
 time properties of $v_{M} (t)$ with similar properties  of ${\cal E}_{M} (t)$
 and $\gamma_{M}(t)$ that have been discussed in Sec. 3.
There is at late times $t \gg T$:
\begin{equation}
\Delta_{M} (t) \equiv \Re\,[v_{M}(t)] \simeq  E_{M} - E_{min}  + \frac{c_{2}}{t^{2}} ....., {\rm for}\;\;t \gg T.
\end{equation}
The late time asymptotic form of $\Im\, [v_{M}(t)] \equiv \Im\,[h_{M}(t)]$ coincides
with the late time form of $\gamma_{M}(t)$ specified by the formula (\ref{E(t)+G(t)}).
\hfill\\

\end{document}